\documentclass[journal,draftcls,onecolumn,12pt,twoside]{IEEEtran}
\usepackage{graphicx,amssymb,amsmath}
\usepackage[noadjust]{cite}%cite
\usepackage{setspace}               % set line space
\usepackage{stfloats}

\usepackage{amsthm}
\usepackage{amsmath}
\usepackage{flushend}
\usepackage{cases,subeqnarray}
\usepackage{bm,multirow,bigstrut}
\usepackage{textcomp}
\usepackage{latexsym,bm}
\usepackage{booktabs,changebar}
\usepackage{xcolor}
\usepackage{mathtools}
\usepackage{dsfont}
\usepackage{extarrows}
\usepackage{mathrsfs}
\usepackage{cite}
\usepackage{bm}
\usepackage{cleveref}
\usepackage{multicol}       % table
\usepackage{multirow}       % table
\usepackage{array}          % table
\definecolor{crimson}{RGB}{192,0,0}         % color crimson
\definecolor{navy}{RGB}{47,85,151}         % color crimson

\makeatletter                               % algorithm
\newif\if@restonecol
\makeatother

\makeatletter
\newif\if@restonecol
\makeatother

\usepackage[linesnumbered,ruled,vlined]{algorithm2e}
\usepackage{algpseudocode}
\usepackage{amsmath}
\theoremstyle{plain}

\theoremstyle{plain}

\def\Htran{\mbox{\tiny $\mathrm{H}$}}

\begin{document}

%----------------------------title&author&thanks----------------------------
\title{Improving Fairness for Cell-Free Massive MIMO Through Interference-Aware Massive Access}
\author{Shuaifei~Chen,~\IEEEmembership{Graduate Student Member,~IEEE}, Jiayi~Zhang,~\IEEEmembership{Senior Member,~IEEE}, Emil Bj{\"o}rnson,~\IEEEmembership{Fellow,~IEEE}, and Bo~Ai,~\IEEEmembership{Fellow,~IEEE}
\thanks{S. Chen, J. Zhang, and B. Ai are with the School of Electronic and Information Engineering and the Frontiers Science Center for Smart Highspeed Railway System, Beijing Jiaotong University, Beijing 100044, China. (e-mail: \{shuaifeichen, jiayizhang, boai\}@bjtu.edu.cn).}
\thanks{E. Bj{\"o}rnson is with the Department of Computer Science, KTH Royal Institute of Technology, SE-16440 Kista, Sweden (e-mail: emilbjo@kth.se).}
}

\maketitle

%----------------------------abstract----------------------------
\begin{abstract}
Cell-free massive multiple-input multiple-output (CF mMIMO) provides good interference management by coordinating {many more access points (APs) than user equipments (UEs).}
It becomes challenging to determine which APs should serve which UEs with which pilots when the number of UEs approximates the number of APs and far exceeds the number of pilots. Compared to the previous work, a better compromise between spectral efficiency (SE) and implementation simplicity is needed in such massive access scenarios.
{This paper proposes an interference-aware massive access (IAMA) scheme realizing joint AP-UE association and pilot assignment for CF mMIMO by exploiting the large-scale interference features. We propose an interference-aware reward as a novel performance metric and use it to develop two iterative algorithms to optimize the association and pilot assignment.
The numerical results show a prominent advantage of our IAMA scheme over the benchmark schemes in terms of the user fairness and the average SE.}

\end{abstract}

%%----------------------------keywords----------------------------
%
%%\newpage
%
\IEEEpeerreviewmaketitle

\section{Introduction}\label{sec:intro}

Cell-free massive multiple-input multiple-output (CF mMIMO) is recognized as a promising paradigm for the sixth-generation (6G) networks \cite{zhang2019multiple}.
The core idea is to coordinate a large number of distributed access points (APs) with a central processing unit (CPU) to provide an almost uniform service quality for the user equipments (UEs) in the coverage area \cite{cellfreebook,chen2022survey}.
However, the enormous wireless devices collaborating not only improves the network throughput, but also introduces a huge challenge on the interference management, especially in the massive access scenarios where the pilot reuse ratio is high.

User access comprises {\it pilot assignment} and {\it AP-UE association}.
The former assigns each UE a pilot to acquire the channel state information (CSI) and the latter associates each UE with at least one AP to perform coherent transmission.
{User access becomes challenging as the network gets more crowded since the increasing UE density implies inevitable pilot reuse among the UEs, which causes substantial inter-user interference, namely {\it pilot contamination} \cite{bjornson2017massive}.}
Also, a high UE density aggravates the competition among UEs for accessing their associated APs, which motivates more delicate AP-UE association schemes.
Moreover, user access should maintain the signal processing operated at each AP under limited complexity and resource requirements to make the system {\it scalable} \cite{bjornson2020scalable}.
The simplest but naive scheme is to assign the pilots and APs at random \cite{ngo2017cell}.
There exist greedy schemes that refine the spectral efficiency (SE) of the weakest UE iteratively \cite{ngo2017cell}, {but cannot guarantee convergence to the global optimal pilot assignment results.}
The graph-based schemes formulate the assignment problems as graph problems (such as the graph coloring \cite{liu2020graph}, the weighted matching \cite{buzzi2020pilot}, and the Max $k$-Cut \cite{zeng2021pilot}), and solve them with the corresponding algorithms.
Scalable schemes are proposed in \cite{bjornson2020scalable} and \cite{chen2020structured}, whereof the former performs joint AP-UE association and pilot assignment, and the latter clusters UEs such that the UEs in the same cluster share the same pilot.

{In a wireless network, being aware of the interference features is critical to the transmission design.
This can be characterized by the treating-interference-as-noise (TIN) optimality conditions from an information-theoretic perspective \cite{geng2015optimality,geng2016optimality,yi2016optimality}, which reflect the interference relationship between an intended link and the two {\it most} influential interfering links corresponding to the intended UE and AP, respectively (as illustrated in Fig.~\ref{fig:system}).}
Especially, \cite{chen2022treating} investigated the probability that the TIN conditions hold in a CF mMIMO system using stochastic geometry, which is directly related to the interference relationships between the APs and the UEs.
Due to the implementation simplicity, robustness to channel uncertainty, and good characterization of interference, {the TIN optimality conditions are exploited for designing interference-aware schemes for scheduling in cellular systems in \cite{bacha2019treating} and power control which further improves the SE in \cite{geng2016optimality}.}

{Motivated by the discussion above, we propose an interference-aware massive access (IAMA) scheme for scalable CF mMIMO, where the interference features are transformed into a useful performance metric for scheme design.
Two assignment algorithms are developed for joint AP-UE association and pilot assignment to maximize the user fairness or the average SE.
The SE improvements achieved by our IAMA scheme are demonstrated by the numerical results.}

{\bf \emph {Notation}}:
Boldface lowercase letters, $\bf x$, denote column vectors, boldface uppercase letters, $\bf X$, denote matrices, and calligraphic uppercase letters, $\cal A$, denote sets.
Superscript $^{\Htran}$ denotes the conjugate transpose.
The $n \!\times \! n$ identity matrix and zero matrix are ${\bf I }_n$ and ${\bf 0}_n$, respectively.
${\cal N}_{\mathbb C}\left({{\bf 0},{\bf R}}\right)$ denotes the multi-variate circularly symmetric complex Gaussian distribution with correlation $\bf R$.
${\mathbb E}\left\{ \cdot \right\}$ denotes the expected value.

\section{CF mMIMO System Model}\label{sec:system}

\begin{figure}[t!]
\centering
\includegraphics[scale=1]{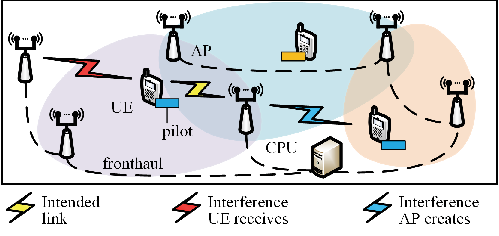}
\caption{Interference relationships in scalable CF mMIMO systems.
\label{fig:system}}
\end{figure}

We consider a CF mMIMO system consisting of $K$ single-antenna UEs and $L$ APs, each equipped with $N$ antennas.
As illustrated in Fig.~\ref{fig:system}, the APs are connected via fronthaul connections to a CPU, which is responsible for coordinating and processing the signals of all UEs.
The user-centric CF architecture is adopted \cite{bjornson2020scalable}, where each UE is associated with a subset of the APs.
This procedure is elaborated in Section \ref{sec:access}.
For now, we let ${\cal M}_k \subset \left\{{ 1,\ldots,L }\right\}$ denote the subset of APs associated with UE $k$ and {let $ {a}_{kl}= 1$ if $l \in {\cal M}_k$ and ${a}_{kl}= {0}$ otherwise, $\forall k,l$.}

We adopt the standard block fading model where the channel between UE $k$ and AP $l$, denoted by ${\bf h}_{kl} \in {\mathbb C}^N$, is constant in time-frequency coherence blocks of $\tau_c$ channel uses \cite{bjornson2017massive}.
In each block, the channels are assumed to be subject to spatially correlated Rayleigh fading, i.e.,
$%\begin{equation}
  {\bf h}_{kl} \sim {\cal N}_{\mathbb C}({\bf 0}, {\bf R}_{kl})
$%\end{equation}
, where ${\bf{R}}_{kl} \in {\mathbb C}^{N \times N}$ is the spatial correlation matrix and {$\beta_{kl} \triangleq {\rm tr}(\frac{{\bf R}_{kl}}{N})$ is the large-scale fading coefficient (LSFC)} that describes pathloss and shadowing.
{We assume that AP $l$ knows the correlation matrices $\{{\bf R}_{kl} : k = 1,\ldots,K\}$, which represent the long-term channel statistics.
These correlation matrices can be accurately estimated using classical methods \cite{bjornson2017massive}.
We consider the downlink operation}, where each block dedicates $\tau_p$ channel uses for pilots and the remaining $\tau_c -\tau_p$ channel uses for payload data.

\subsection{Data Transmission and Spectral Efficiency}

During the channel estimation, $\tau_p \!<\! K$ holds in massive access scenarios due to the coherence block length limitation caused by natural channel variations in the time and frequency domain.
We adopt that the pilots are selected from a pool of $\tau_p$ orthogonal sequences, and thus, some UEs have to share the same pilot.
We let ${\cal P}_{t}$ denote the set of UEs sharing pilot $t$ and refer to these UEs as {\it co-pilot UEs}.
The pilot assignment is elaborated in Section \ref{sec:access}.
For now, we denote by $t_k \in \{1,\ldots,\tau_p\}$ the index of the pilot assigned to UE $k$.
When the UEs in ${\cal P}_{t_k}$ transmit pilot $t_k$, the pilot signal ${\bf{y}}_{{t_k}l}^{\rm{p}} \in {{\mathbb C}^N}$ received at AP $l$ is \cite[Sec. 3]{bjornson2017massive}
\begin{equation}\label{eq:received pilot signal at AP}
  {\bf y}_{{t_k}l}^{\rm{p}} = \sum\nolimits_{i \in {{\cal P}_{t_k}}} {\sqrt {{\tau _p}{\rho _{\rm{p}}}} {{\bf{h}}_{il}}}  + {{\bf{n}}_{{t_k}l}},
\end{equation}
where $\rho_{\rm p}$ represents the pilot transmit power and ${{\bf{n}}_{{t_k}l}} \!\sim\! {{\cal N}_{\mathbb{C}}}\left( {{\bf{0}},{\sigma ^2}{{\bf{I}}_N}} \right)$ is the thermal noise.
The \emph{minimum mean-squared-error (MMSE)} estimate of ${\bf h}_{kl}$ is \cite[Sec. 3]{bjornson2017massive}
\begin{equation}\label{eq:mmse estimate}
  {{{\bf{\widehat h}}}_{kl}} \!=\! \sqrt {{\tau _p}{\rho _{\rm{p}}}} {{\bf{R}}_{kl}}{\bf{\Psi }}_{{t_k}l}^{ - 1}{\bf y}_{{t_k}l}^{\rm{p}} \sim {\cal N}_{\mathbb C}\left({ {\bf 0},{\tau _{\rm{p}}}{\rho _{\rm{p}}}{{\bf{R}}_{kl}}{\bf{\Psi }}_{{t_k}l}^{ - 1}{{\bf{R}}_{kl}} }\right),
\end{equation}
where ${{\bf{\Psi }}_{{t_k}l}} = \sum\nolimits_{i \in {{\cal P}_{t_k}}}  {{\tau _p}{\rho _{\rm{p}}}{{\bf{R}}_{il}}}  + {\sigma ^2}{{\bf{I}}_N}$
is the correlation matrix of ${\bf y}_{{t_k}l}^{\rm{p}}$ in \eqref{eq:received pilot signal at AP}.

Let ${{\bf{w}}_{kl}} \!=\! {{{\bf{\bar w}}}_{kl}}/{\sqrt{{\mathbb{E}}\{ {\| {{{{\bf{\bar w}}}_{kl}}} \|^2}\}}} $ denote the normalized precoder that AP $l$ selects for transmission to UE $k$ such that ${\mathbb E}\{\|{\bf{w}}_{kl}\|^2\} \!= \! 1$.
Then the received signal at UE $k$ is
\begin{equation}\label{eq:received data signal at UE}
  {y_k^{{\rm{dl}}} = \sum\nolimits_{l = 1}^L {{\bf{h}}_{kl}^{\Htran}\sum\nolimits_{i = 1}^K {{a_{il}}{{\bf{w}}_{il}}{s_i}} }  + {n_k},}
\end{equation}
where $s_k \in {\mathbb C}$ is the independent unit-power payload signal intended for UE $k$, $\rho_{il}\ge 0$ is the transmit power that AP $l$ assigns to UE $i$, and $n_k \sim {\cal N}_{\mathbb C}(0,\sigma^2)$ is the receiver noise.
The total transmission power of each AP is upper bounded by the maximum power $\rho_{\rm dl}$.

We employ the widely used \emph{hardening bound} \cite[Th. 4.6]{bjornson2017massive},
\begin{equation}\label{eq:DL_SE}
{\sf SE}_k = \left( {1 - {\tau_p}/{\tau_c}} \right){\log _2}\left( {1 + {\sf SINR}_k} \right),
\end{equation}
to compute the achievable downlink SE, where the signal-to-interference-and-noise ratio (SINR) is given by
\begin{equation}\label{eq:sinr}
{{\sf SINR}_k \!=\! \small \frac{ \left| {\mathbb E}\left\{ \sum\limits_{l=1}^L a_{kl} {\bf h}_{kl}^{\Htran} {\bf w}_{kl}  \right\} \right|^2}
{\sum\limits_{i=1}^K \left| {\mathbb E}\left\{ \sum\limits_{l=1}^L a_{il} {\bf h}_{kl}^{\Htran} {\bf w}_{il}  \right\} \right|^2 - \left| {\mathbb E}\left\{ \sum\limits_{l=1}^L a_{kl} {\bf h}_{kl}^{\Htran} {\bf w}_{kl}  \right\} \right|^2}.}
\end{equation}
The SE expression in \eqref{eq:DL_SE} holds for any scalable precoding scheme, e.g., the local partial MMSE (LP-MMSE) precoding \cite{bjornson2020scalable} or the classical maximum ratio (MR) precoding \cite{bjornson2017massive}.

{\subsection{Interference-Aware Rewards (IARs) for CF mMIMO}}

The TIN optimality conditions indicate when the boundary of the capacity region is approached within a constant gap and power control is essential to do that \cite{geng2015optimality}.
In cellular networks, interference-aware scheduling schemes are designed based on the following TIN condition \cite{bacha2019treating}:
\begin{equation}\label{eq:tin_cell}
\kappa {\sf{SNR}} ^ \mu \ge \max {\sf{INR}_{{\rm{ap}}}} \cdot \max {\sf{INR}_{{\rm{ue}}}},
\end{equation}
where ${\sf{SNR}}$ denotes the signal-to-noise ratio (SNR) of the intended link, ${\sf{INR}_{{\rm{ap}}}}$ denotes the interference-to-noise ratio (INR) of the link between the intended AP and the interfering UE and ${\sf{INR}_{{\rm{ue}}}}$ denotes the INR of the link between the intended UE and the interfering AP.
{In \eqref{eq:tin_cell}, ${\sf SNR}$, ${\sf INR}_{\rm ap}$, and ${\sf INR}_{\rm ue}$ rely on statistical knowledge, which is averaged concerning the fast fading.}
Parameters $\kappa \ge 1$ and $1\le \mu \le 2$ are introduced in \cite{bacha2019treating} for system optimization.

The condition in \eqref{eq:tin_cell} will not directly be applied to our considered CF user access since a UE is served by multiple APs.
In our case, the interference comes from the imperfect CSI caused by the pilot reuse, which will both reduce the channel estimation quality and make it harder to suppress interference among the co-pilot UEs.
Similar to \cite{chen2022treating}, we denote by
\begin{equation}
{{\cal S}}_{t,k}^{\rm ue} = {\cal P}_{t} \setminus \{k\}
\end{equation}
the set of UEs sharing pilot $t$ except UE $k$, which are referred to as the {\it interfering UEs} of UE $k$ when UE $k$ is assigned with pilot $t$.
Also, we denote by
\begin{equation}
{{\cal S}}_{t,k}^{{\rm{ap}}} = \bigcup\nolimits_{i \in {{\cal S}}_{t,k}^{\rm ue}} {\cal M}_i \setminus {\cal M}_k
\end{equation}
the set of the {\it interfering APs} of UE $k$ when UE $k$ is assigned with pilot $t$.
Unlike the scheduling in \cite{bacha2019treating} where only the links fulfilling the TIN condition in \eqref{eq:tin_cell} are established, {we propose a novel performance metric tailored for CF mMIMO, as
\begin{equation}\label{eq:tin_reward}
 {\sf iar}_{kl}^t \triangleq
\kappa {\left( {{\beta}_{kl}} \right)^\mu } - {\max _{i \in {{\cal S}}_{t,k}^{{\rm{ue}}}}}{\beta_{il}} \cdot {\max _{j \in {{\cal S}}_{t,k}^{{\rm{ap}}}}}{\beta_{kj}}, \ \forall k,l,t,
\end{equation}
which reflects the strength of an intended link compared to the interference by only utilizing the LSFCs.
We refer to ${\sf iar}_{kl}^t$ defined in \eqref{eq:tin_reward} as the {\it interference-aware reward (IAR)} of a tuple $(k,l,t)$, which corresponds to the strength of the link between AP $l$ and UE $k$ when pilot $t$ is used.}

{
\section{Interference-Aware Massive Access (IAMA)}\label{sec:access}
In this section, we propose an IAMA scheme that uses the proposed IARs in \eqref{eq:tin_reward} with the goal of improving the SE of the majority of UEs, e.g., the $90\%$-likely SE, which is a well-used performance criterion representing the SE that can be provided to $90\%$ of all UEs \cite{cellfreebook}.
Since a new UE needs to be assigned a pilot to perform coherent transmission with its associated APs when it accesses the network, two constraints should be met during the AP-UE-pilot association:
\begin{itemize}
  \item Each UE is associated with at least one AP to not being dropped from service inadvertently;
  \item Each AP serves at most one UE per pilot to avoid causing substantial pilot contamination.
\end{itemize}
As illustrated in Fig.~\ref{fig:procedure}, our proposed IAMA scheme operates through three steps: 1) Master AP (mAP) selection; 2) pilot assignment; and 3) further AP-UE association. Since only the LSFCs among the APs and the UEs are employed at the CPU, the IAMA scheme works for many coherence blocks.

Further details on above steps are provided later in this section.
Before that, recall that we use $a_{kl} \in \{0,1\}$, $\forall k,l$, to indicate the AP-UE association, we introduce another binary notation $b_{kl} \in \{0,1\}$, $\forall k,l$, to ensure the association algorithm convergence.
More precisely, during the access procedure, $b_{kl} \!=\! 1$ prevents the association between AP $l$ and UE $k$ from being considered again, and $b_{kl} \!=\! 0$ otherwise.

\begin{figure}[t!]
\centering
\includegraphics[scale=0.85]{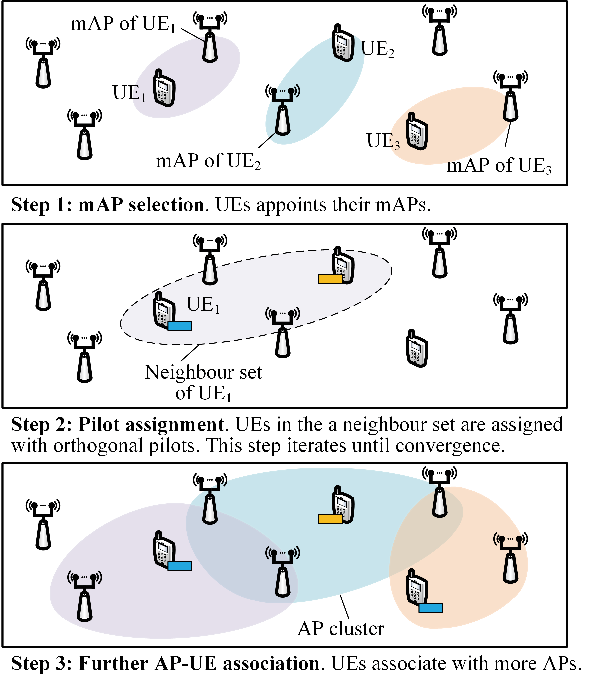}
\caption{The proposed IAMA scheme for joint pilot assignment and AP-UE association operates through three main steps.
\label{fig:procedure}}
\end{figure}

\subsection{mAP Selection}

Similar to \cite{bjornson2020scalable}, each UE first appoints a {\it mAP} assisting in the following pilot assignment and AP-UE association.
One well-used approach is to let each UE select the AP with largest LSFC as its mAP, however, with the risk that an AP is selected as the mAP by more than $\tau_p$ UEs.
This motivates us to develop the Multiple-UEs Single-AP-each (MUSA) algorithm to assign each UE to one mAP while each AP is assigned to at most $\tau_p$ UEs as their common mAP.
The MUSA algorithm tries to maximize the total LSFC of all associated AP-UE pairs, i.e., $\sum\nolimits_{k,l} {\beta}_{kl}\cdot a_{kl}$, and consists of the following steps:
\begin{enumerate}
  \item Each UE initially appoints the AP that it has the largest LSFC to as its {\it candidate} mAP.
  \item Find the {\it overburdened} APs selected by more than $\tau_p$ UEs and include them in cluster ${\cal C}_{\rm ap} \!=\! \{l: \!\sum_{k} a_{kl} \!>\! \tau_p\}$.
  \item For each AP $l \in {\cal C}_{\rm ap}$, include the UEs selecting AP $l$ as the mAP in cluster ${\cal C}_{\rm ue} \!=\! \{k: a_{kl} \!=\! 1, l \in {\cal C}_{\rm ap} \}$.\\
      For each UE $k \!\in\! {\cal C}_{\rm ue}$, compute the {\it LSFC loss} as ${\Delta_k} \!\triangleq\! {\beta_{kl}} \!-\! {\beta_{k\ell_k}}$, where AP $\ell_k \!=\! \arg {\max _{j \ne l, b_{kj} \ne 1}}{{\beta}_{kj}}$ is the {\it alternative} mAP having the largest LSFC with UE $k$ except AP $l$.\\
      Find UE $i = \arg {\min _{k \in {\cal C}_{\rm ue}}}{{\Delta}_{k}}$ with the smallest LSFC loss, replace the mAP of UE $i$ by its AP $\ell_i$, and mark the association between AP $l$ and UE $i$ by letting $b_{il}=1$.
  \item Repeat steps 2) and 3) until ${\cal C}_{\rm ap} = \emptyset$ or $\sum_{k} b_{kl} = K, \forall l \in {\cal C}_{\rm ap}$.
\end{enumerate}
The pseudo-code of MUSA is summarized in Algorithm \ref{algo:musa}.

\begin{algorithm}[t!]
\label{algo:musa}
{
%\doublespacing
\caption{MUSA Assignment}
\KwIn{$\{\beta_{kl}:\forall k,l\}$, $\tau_p$}

\KwOut{$\{a_{kl}:\forall k,l\}$}

{\bf Initiation:} $a_{kl} = 0, b_{kl} = 0,\forall k,l$;\\

\For{{\rm UE} $k =1,\ldots,K$}
{
Select AP $\ell = \arg {\max _{l}}{\beta_{kl}}$ as its {\it candidate} mAP and let $a_{k\ell} \leftarrow 1$;\\
}
Include the {\it overburdened} APs in $
{\cal C}_{\rm ap} \!=\! \{l: \sum\nolimits_{k} a_{kl} \!>\! \tau_p \}
$;\\
\While{${\cal C}_{\rm ap} \ne \emptyset$ and $\sum\nolimits_{k} b_{kl} < K, \exists l\in {\cal C}_{\rm ap}$}
{
\For{{\rm overburdened AP} $l \in {\cal C}_{\rm ap}$}
{
Include the UEs selecting AP $l$ in ${\cal C}_{\rm ue} \!=\! \{k: a_{kl} \!=\! 1, l \in {\cal C}_{\rm ap} \}$;\\
Let ${\Delta}_k = 0, k \in {\cal C}_{\rm ue}$;\\
\For{{\rm UE} $k \in {\cal C}_{\rm ue}$}
{
Find {\!\it alternative\!} mAP $\ell_k \!=\! \arg \!{\max _{j \ne l, b_{kj} \ne 1}}{\beta_{kj}}\!$ and compute LSFC loss ${{\Delta}_{k}} = {\beta_{kl}} - {\beta_{k\ell_k}}$;\\
}
Find UE $i = \arg {\min _{k \in {\cal C}_{\rm ue}}}{{\Delta}_{k}}$, replace its mAP as AP $\ell_i$ by $a_{il} \leftarrow 0$ and $a_{i\ell_{i}} \leftarrow 1$,
and mark the association between AP $l$ and UE $i$ by $b_{il} \leftarrow 1$;\\
Update cluster $
{\cal C}_{\rm ap} \!=\! \{l: \sum\nolimits_{k} a_{kl} \!>\! \tau_p \}$;
}
}
}
\end{algorithm}

\subsection{Pilot Assignment}

To reduce the pilot contamination, the UEs prefer to be assigned pilots that are orthogonal to their neighbouring UEs.
With this consideration in mind, we propose the following IAR-based pilot assignment scheme:
\begin{enumerate}
  \item Each UE selects a pilot from the $\tau_p$ pilots at random.
  \item Consider a generic UE $k$ and include it and the $\tau_p - 1$ neighbouring UEs having the largest LSFCs with the mAP of UE $k$ (i.e., AP $l_k$) in UE $k$'s {\it neighbour} set ${\cal N}_k$, where $|{\cal N}_k| = \tau_p$.\\
      By using \eqref{eq:tin_reward}, compute all {\it potential} IARs of the tuples $(i,l_k,t)$, i.e., $\{{\sf iar}_{i l_k}^{t}\!\!: t \!=\! 1,\ldots,\tau_p, i\in {\cal N}_k\}$.
  \item Update the pilot assignment in ${\cal N}_k$ by performing the MUSA assignment in Algorithm \ref{algo:musa}, where the inputs are replaced by $\{{\sf iar}_{i l_k}^{t}\}$ and integer $1$, and the output indicates which pilot is assigned to which UE in ${\cal N}_k$.
  \item Repeat steps 2) and 3) until the maximum number of allowed iterations is reached or convergence, measured by the change in the sum IAR $\sum_{i=1}^{K} {\sf iar}_{i l_i}^{t_i}$.
\end{enumerate}

Step 3) implies that MUSA assigns each UE one pilot while each pilot is assigned to at most one UE, based on the IARs $\{{\sf iar}_{i l_k}^{t}\!\!: t \!=\! 1,\ldots,\tau_p, i\in {\cal N}_k\}$, $\forall k$.

\subsection{Further AP-UE Association}

Given the assigned mAP and pilot, each UE prefers to access more serving APs to improve the diversity gain. This motivates our Multiple-UEs Multiple-APs-each (MUMA) algorithm that optimizes the AP-UE association, based on the {\it potential} IARs of the tuples $(k,l,t_k)$, i.e.,
\begin{equation}\label{eq:tin_reward1}
\{{\sf iar}_{k l}^{t_k}\!\!: k \!=\! 1,\ldots,K, l \!=\! 1,\ldots,L\}.
\end{equation}
Recall that although our IAMA scheme aims to improve the SE of most UEs, an option for improving the average SE is also provided.
More precisely, the MUMA algorithm either maximizes the user fairness as $\max_{\{a_{kl}\}} \min_k \Sigma_k$, where $\Sigma_k \triangleq \sum\nolimits_{l} {\sf iar}_{k l}^{t_k} a_{kl}$ denotes the per-UE sum IAR, or maximizes the total sum IAR as $\max_{\{a_{kl}\}} \sum\nolimits_{k,l} {\sf iar}_{k l}^{t_k} a_{kl}$.
We use the association between the UEs and their mAPs to initialize $\{b_{kl},\forall k,l\}$.
The MUMA algorithm operates as follows:
\begin{enumerate}
  \item By using \eqref{eq:tin_reward}, compute all potential IARs in \eqref{eq:tin_reward1}.
  \item Each AP associates $\tau_p$ UEs with the largest IARs. \\
  If the goal is $\max_{\{a_{kl}\}} \sum\nolimits_{k,l} {\sf iar}_{k l}^{t_k} a_{kl}$, stop algorithm and return $\{a_{kl}\}$; otherwise, continue.
  \item Find the {\it weakest} UE
  $
  k' = \arg {\min _{k}} {{\Sigma}}_k
  $ with the smallest per-UE sum IAR.
  \item Find the {\it closest} AP
  $
  l' \!=\! \arg {\max _{l,b_{k'l} \ne 1}} {\sf iar}_{k'l}^{t_{k'}}
  $
  for UE $k'$, satisfying $b_{k'l} \ne 1$ with the largest IAR.
  \item Find the {\it most distant} UE
  $
  k^\ast = \arg {\min _{k,b_{kl'} \ne 1}} {\sf iar}_{kl'}^{t_{k}}
  $
  for AP $l'$, satisfying $b_{kl'} \ne 1$ with the smallest IAR.
  \item Mark the association between AP $l'\!$ and UE $k'\!$ by $b_{k'l'} \!\leftarrow\! 1$.
   If UE $k'\!$ still has a smaller per-UE sum IAR than UE $k^*\!$ after taking AP $l'$ from UE $k^*\!$, then UE $k'\!$ takes AP $l'\!$ from UE $k^*\!$; otherwise, keep status quo.
  \item Repeat step from 3) to 6) until $\sum\nolimits_{l} b_{kl} = L$, $\exists k$.
\end{enumerate}
The pseudo-code of MUMA is summarized in Algorithm \ref{algo:muma}.
%, where we use the more generalized input notations $\bf W$ and $n$ for the same reason as in Algorithm \ref{algo:musa}.

\begin{algorithm}[t!]
\label{algo:muma}
%\doublespacing
{
\caption{MUMA Assignment}
\KwIn{$\{{\sf iar}_{k l}^{t_k}:\forall k,l\}$, $\{a_{kl}:\forall k,l\}$, $\{b_{k l}= a_{kl}:\forall k,l\}$, $\tau_p$}

\KwOut{$\{a_{kl}:\forall k,l\}$}

\For{{\rm AP} $l = 1,\ldots,L$}
{
Sort $\{{\sf iar}_{1 l}^{t_1},\ldots,{\sf iar}_{K l}^{t_K}\}$ in descending order and include $\tau_p$ UEs with the largest values in cluster ${\cal C}_{\rm ue}$;\\
Associate the UEs in ${\cal C}_{\rm ue}$ by letting $a_{il} = 1, i \in {\cal C}_{\rm ue}$;\\
}
\If{The goal is $\max_{\{a_{kl}\}} \sum\nolimits_{k,l} {\sf iar}_{k l}^{t_k} a_{kl}$}
{
{\bf Return};\\
}
\ElseIf{The goal is $\max_{\{a_{kl}\}} \min_k {{\Sigma}}_k$}
{
\While{$ \sum\nolimits_{l} b_{kl} < L, \forall k$}
{
Compute the per-UE sum IARs ${{\Sigma}}_k, \forall k$;\\
Find UE $k' = \arg {\min _{k}}{{{\Sigma}}_k}$;\\
Find AP
$l' = \arg {\max _{l,b_{k'l} \ne 1}} {\sf iar}_{k'l}^{t_{k'}}$ for UE $k'$;\\
Find UE $k^\ast = \arg {\min _{k,b_{kl'} \ne 1}} {\sf iar}_{kl'}^{t_{k}}$ for AP $l'$;\\
$b_{k'l'} \leftarrow 1$;\\
\If{${{\Sigma}}_{k^\ast} - {\sf iar}_{k^\ast l'}^{t_{k^\ast}} \le {{\Sigma}}_{k'}$}
{
$b_{k^\ast l} \leftarrow 1$, where $l\in \{j: a_{k^\ast j} = 1\}$;\\
{\bf{Continue}};\\
}
$b_{k'l} \leftarrow 1$, where $l\in \{j: a_{k'j} = 1\}$;\\
$a_{k^\ast l'} \leftarrow 0$, $a_{k'l'} \leftarrow 1$;\\
}
}
}
\end{algorithm}
\subsection{Benchmark Schemes and Complexity Analysis}

Three existing user access schemes are considered as benchmarks, they are ``Scalable" with the complexity of ${\cal O}(KL+ K \tau_p +L\tau_p)$ \cite{bjornson2020scalable}, ``Greedy" with the complexity of ${\cal O}(3KL+L)$ \cite{ngo2017cell}, and ``Graph" with the complexity of ${\cal O}(KL+K^2/2+K/2+\tau_p)$ \cite{zeng2021pilot}.
The Hungarian scheme proposed in \cite{buzzi2020pilot} is not considered since Graph offers better SE \cite{zeng2021pilot}.
For fair comparison in the scalable scenario, each AP serves at most $\tau_p$ UEs and allocates its transmit power with the fractional power allocation policy \cite{bjornson2020scalable}.

We consider two use cases of our proposed IAMA scheme: a) ``IARsum" for maximizing the sum SE and b) ``IARmin" for maximizing the user fairness.
The complexity of MUSA depends on computing $\Delta_k$ for UE $k\in {\cal C}_{\rm ue}$ of APs in ${\cal C}_{\rm ap}$, from line 5 to line 12 in Algorithm \ref{algo:musa}, which traverses all entries in $\{\beta_{kl},\forall k,l\}$ at most $K$ times, and, thus, with the complexity of ${\cal O}(K^2 L)$ \cite{jungnickel2005graphs}.
%is approximately the same as the Hungarian algorithm, i.e., ${\cal O}(K^2 L)$ \cite{buzzi2020pilot}.
The complexity of MUMA depends on comparing the IARs of the weak UEs, from line 7 to line 17 in Algorithm \ref{algo:muma}, which traverses all entries in $\{{\sf iar}_{k l}^{t_k}:\forall k,l\}$ at most $L$ times, and, thus, with the complexity of ${\cal O}(KL^2)$ \cite{jungnickel2005graphs}.
To sum up, the total complexity of the IAMA scheme is ${\cal O}(K^2 L + K\tau_p^3 +KL^2)$.
For the considered massive access scenario where $K \!\approx \! L \!\gg\! \tau_p$, the complexity of the IAMA scheme is dominated by ${\cal O}(K^3)$.
}

\begin{figure}[t!]
\centering
\includegraphics[scale=0.63]{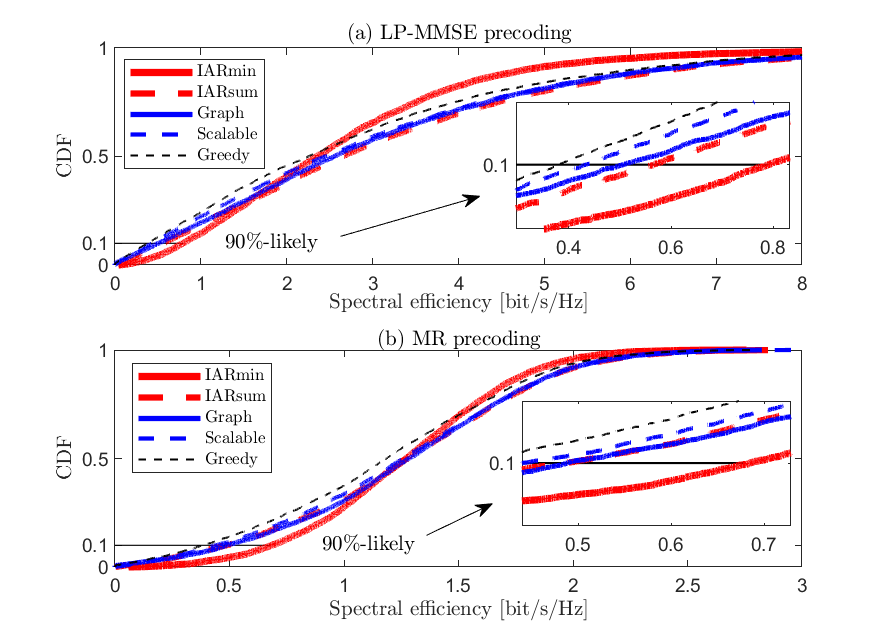}
\caption{Downlink SE per UE with different access schemes ($K=50$).
\label{fig:cdf_se k50}}
\end{figure}

\section{Numerical Results}\label{sec:results}

{In this section, we quantify the SE performance of the proposed IAMA scheme, in which the results regarding LP-MMSE are obtained from Monte Carlo simulation while the results regarding MR are analytically computed by \cite[Cor. 3]{bjornson2020scalable}.}
We consider a $0.5\!\times\! 0.5$ km$^2$ coverage area and use the wrap-around technique to approximate an infinitely massive access scenario, where $L\!=\!50$ APs are deployed at random and each is equipped with a half-wavelength-spaced uniform linear array with $N \!=\!4$ antennas.
Two different numbers of UEs are considered: a) $K\!=\!50$ corresponding to 200 UEs/km$^2$ and b) $K\!=\!100$ corresponding to 400 UEs/km$^2$.
The 3GPP Urban Microcell model is used to compute the large-scale propagation conditions, such as pathloss and shadow fading.
Unless otherwise specified, other system parameters are referred to \cite{bjornson2020scalable,chen2022treating}, which are $\rho_{\rm p} \!=\! 0.1$ W, $\rho_{\rm dl} \!=\! 1$ W, $\sigma^2 \!=\! -94$ dBm, $\tau_p = 5$, $\tau_c = 200$, $\kappa\! =\! 10$, and $\mu \!=\! 1.8$.

\begin{figure}[t!]
\centering
\includegraphics[scale=0.63]{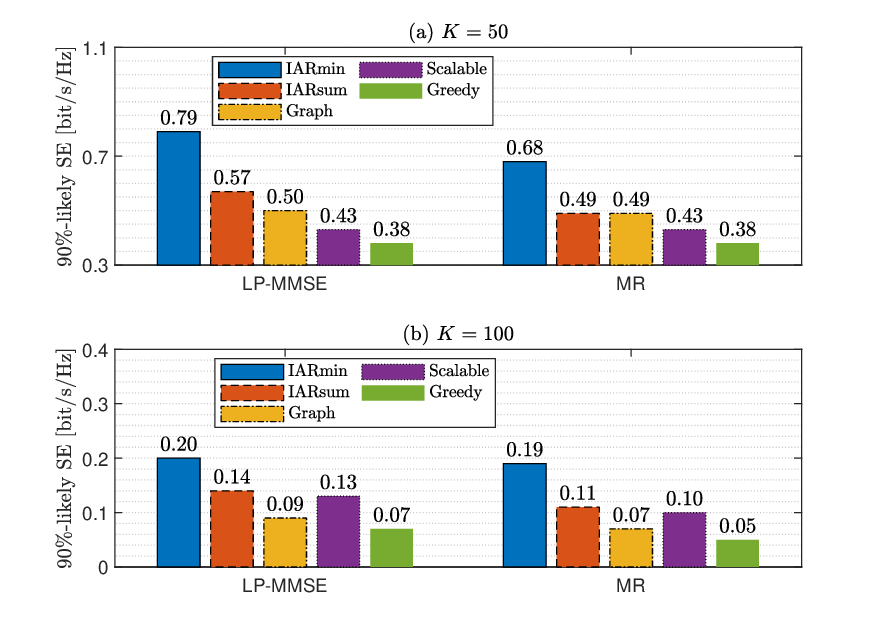}
\caption{90$\%$-likely SE with different numbers of UEs and precoders.
\label{fig:90_se}}
\end{figure}

\begin{figure}[t!]
\centering
\includegraphics[scale=0.63]{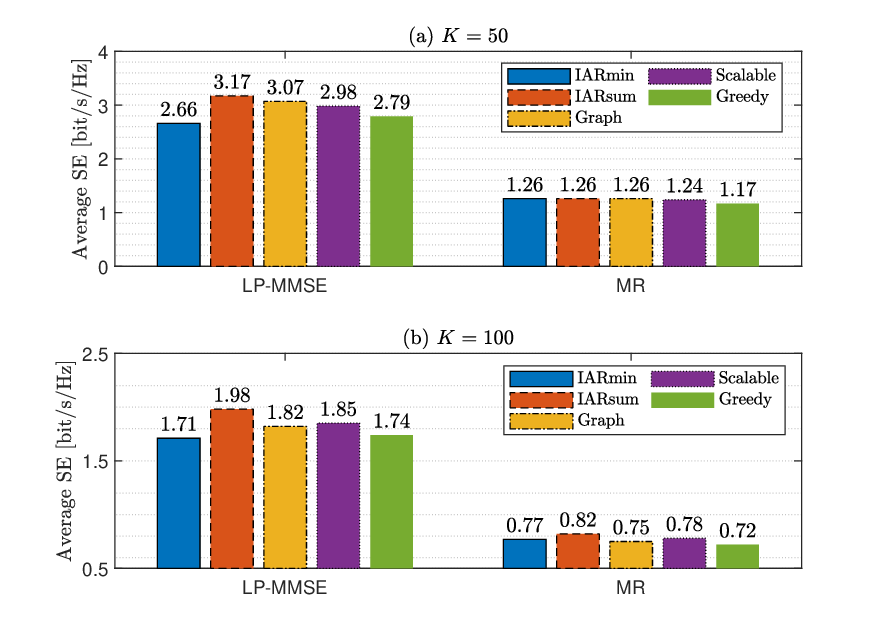}
\caption{Average SE with different numbers of UEs and precoders.
\label{fig:mean se}}
\end{figure}

{
Fig.~\ref{fig:cdf_se k50} shows the cumulative distribution function (CDF) of the downlink SE per UE with $K\!=\!50$, which gives a macro perspective comparison among the schemes.
The most prominent observation is that IARmin significantly outperforms IARsum and other schemes on the user fairness (quantified by the $90\%$-likely SE value) with both precoders.
There are two reasons.
On the one hand, MUSA ensures that each weak UE with poor channel condition accesses at least one AP and satisfies that each AP serves at most one UE per pilot.
On the other hand, MUMA tries to associate the weak UEs with more serving APs which promotes the user fairness.
Without the max-min association readjustment in MUMA (i.e., from line 6 to line 17 in Algorithm \ref{algo:muma}), IARsum falls behind IARmin, but still provides a higher $90\%$-likely SE than the other considered schemes due to the accurate interference characterization of the IARs in \eqref{eq:tin_reward} such that each UE is assigned the best pilot causing the least pilot contamination.

In Fig.~\ref{fig:90_se}, we elaborate the $90\%$-likely SE with different numbers of UEs and precoders.
Fig.~\ref{fig:90_se}(a) quantifies the $90\%$-likely SE in Fig.~\ref{fig:cdf_se k50}, from which we can see that IARmin achieves $58\%$ and $39\%$ higher $90\%$-likely SE than the best benchmark (i.e., Graph) when using LP-MMSE precoding and MR precoding, respectively.
Fig.~\ref{fig:90_se}(b) compares the considered schemes in a denser scenario with $K\!=\!100$.
When compared to Fig.~\ref{fig:90_se}(a), it is clear that the $90\%$-likely SE is deteriorated by the severe pilot contamination, while the advantage of IARmin grows.
More precisely, we observe that IARmin achieves $67\%$ and $90\%$ higher $90\%$-likely SE than the best benchmark (i.e., Scalable) when using LP-MMSE precoding and MR precoding, respectively.

The average SE of the considered schemes is evaluated in Fig.~\ref{fig:mean se}.
Although improving the average SE is not the main goal of our IAMA scheme, IARsum still outperforms the benchmarks in all considered cases.
When comparing Fig.~\ref{fig:mean se}(a) and (b), we can see the advantage of IARsum grows when the number of UEs increases, benefiting from the protection mechanism for the weak UEs of MUSA.
For example, when using LP-MMSE precoding, the average SE of IARsum is slightly higher than the best benchmark Graph with $K=50$ while the improvement compared to the best benchmark Scalable becomes $7\%$ with $K=100$.
IARmin loses the average SE for improving the $90\%$-likely SE, but still slightly outperforms some benchmarks in the MR cases, where the weak UEs rely on accessing more serving APs for avoiding interference.
}

\section{Conclusion}\label{sec:conclusion}

{We proposed an interference-aware massive access scheme for scalable CF mMIMO in this paper.
We proposed an iterative procedure for joint AP-UE association and pilot assignment by exploiting the proposed IARs, where two assignment algorithms are developed to maximize the user fairness or the average SE.
The numerical results showed that our IAMA scheme significantly improved the user fairness compared to the state-of-the-art benchmark schemes, especially when using LP-MMSE precoding in a denser scenario.}

\bibliographystyle{IEEEtran}
\bibliography{IEEEabrv,ref}

\end{document}